\documentclass[smallextended]{svjour3}      
\usepackage{graphicx}
\usepackage{hyperref}
\usepackage{color}
\usepackage{amsmath}
\usepackage{amsfonts}
\usepackage{amssymb}
\usepackage{array}
\usepackage{multirow}
\begin{document}
\title{Quantum physics, digital computers, and life from a holistic perspective}

\author{George F R Ellis} 
\institute{George F R Ellis \at
          Mathematics Department, University of Cape Town \\
             Rondebosch, Cape Town 7701, South Africa\\
               \email{george.ellis@uct.ac.za} 
 \at The New Institute, Warburgstrasse 8, \\
Hamburg 20354, Germany\\
\email{george.ellis@thenew.institute}}

\maketitle
\begin{abstract}
\textit{Quantum physics is a linear theory, so it is somewhat puzzling that it can underlie very complex systems such as digital computers and life.  This paper investigates how this is possible. Physically, such complex systems are necessarily modular hierarchical structures, with a number of key features. Firstly, they cannot be described by a single wave function: only local wave functions can exist, rather than a single wave function for a living cell, a cat, or a brain. Secondly,  the quantum to classical transition is characterised by contextual wave-function collapse shaped by macroscopic elements that can be described classically. Thirdly, downward causation occurs in the physical hierarchy in two key ways: by the downward influence of time dependent constraints, and by creation, modification, or deletion of lower level elements. 
Fourthly, there are also logical modular hierarchical structures supported by the physical ones, such as algorithms and computer programs, They are able to support arbitrary logical operations, which can influence physical outcomes as in computer aided design and 3-d printing. Finally, complex systems are necessarily open systems, with heat baths playing a key role in their dynamics and providing local arrows of time that agree with the cosmological direction of time that is established by the evolution of the universe. }  
\end{abstract}

\section{Introduction}
\label{sec:Intro}
A crucial feature of life is the way function, purpose, and agency arise in biology \cite{Ball_2023,Mitchell_2023,Noble_Noble_2023}
despite none of these attributes occurring at the underlying physics levels \cite{Hartwell_1999,Ellis_2016}. In particular these features are not apparent in the quantum dynamics  \cite{Landshoff_et_al,Laughlin_Pines,Isham_2001} underlying the functioning of life.  Furthermore digital computers are able to carry out abstract causation \cite{Dasgupta_2016}, involving logical branching deployed in order to attain some desired outcome. This again is based ultimately in the underlying quantum physics \cite{Ellis_Drossel_computers}; but quantum physical processes do not by themselves involve logical branching such as is key to digital computation. 
This paper looks at the way the underlying quantum physics relates to these issues and enables such real-world emergence to occur. 

A first key feature is that quantum theory, via the Schr\"{o}dinger equation \cite{Landshoff_et_al,Laughlin_Pines,Isham_2001}, is linear in the wave function $|\Psi\rangle$ \cite{Ellis_2012}, but the real universe is highly non-linear; consequently only local wave functions can exist \cite{Ellis_2023}. It is this that for example allows feedback control loops and homeostasis to exist.

A second key feature is the downward causation that underlies emergence of complexity in modular hierarchical systems \cite{Ellis_2016}. There are two  ways this occurs \cite{Ellis_2023a}: via time dependent constraints, which is where the  potential term in the Hamiltonian enters; and via creation or deletion of lower level elements, which is where quantum field theory enters. A crucial feature of downward causation is multiple realizability of higher levels at lower levels; and this applies both to physical hierarchies, and the logical hierarchies they support.

Thirdly, downwards causation from the environment occurs because any complex system is an open system coupled to heat baths that play a key role in classical and quantum mechanics \cite{Drossel_2014}, and specifically in contextual wave function collapse  \cite{CWC}. This is how local arrows of time arise out of the cosmological direction of time associated with the expansion of the universe.

Finally, downward causation associated with mental causation is a product of the human mind, with  choices made crucially shaped by the values of those making decisions. This is central to the functioning of digital computers \cite{Ellis_Drossel_computers}, as is apparent in the way social media influences the world around us. \\

\subsection{Symbol structures and abstract causation} 

A key point to make at the start is that (\cite{Dasgupta_2016}:11,12):
\begin{quote}
    ``As regards digital computers, in its most fundamental, the stuff of computers is systems of symbols, forming \textit{symbol structures}- that is entities that stand for, denote, or represent other entities. \textit{Computing is symbol processing}.''
\end{quote}
Then because we are a symbolic species \cite{Deacon}, essentially the same is true for the human brain. \textit{Inter alia}, amongst its capacities is logical thinking:
\begin{quote}
    As regards the human brain, in its most fundamental, the stuff of thought is \textit{symbol structures.  Thinking is symbol processing}.
\end{quote}
This is the basis of abstract causation, where symbols cause changes in the physical world - as happens all the time in society (think traffic lights).

\paragraph{Abstract Entities have causal powers}
This is clear in digital computers \cite{Dasgupta_2016,Ellis_Drossel_computers}, where (1) algorithms, (2) computer programs, and (3) data - all abstract entities - have causal powers because they alter physical outcomes in a real-world social context. 
Digital computers are at their heart symbol processing systems. As expressed in (\cite{Dasgupta_2016}:23),
\begin{quote}
\textit{Some computational artefacts are entirely abstract: they not only process symbol structures, they} themselves \textit{are symbol structures and are devoid of any physicality, though they may be made visible by physical media such as marks on paper or computer screens. so physico-chemical laws do not apply to them. They neither occupy physical space nor do they consume physical time}.
\end{quote}
However by the processes detailed in this paper and in \cite{Dasgupta_2016,Ellis_Drossel_computers}, these abstract entities have causal powers via what Dasgupta  refers to as \textit{Liminal Structures} where the abstract becomes physical and can result in images on a screen, marks on paper, 3-d printing of artefacts, automatic landing of an aircraft, and so on. 
Similar remarks apply to the brain.

\subsection{This paper}
The following sections look at, Hierarchies of structure and causation (\S\ref{sec:2}); Quantum linearity, local wave functions, and wave function collapse (\S\ref{sec:3}); Quantum Theory and time dependent constraints (\S\ref{sec:4}); Quantum Field Theory: particle creation and annihilation (\S\ref{sec:5}); Open systems and the arrow of time (\S\ref{sec:6}); and, Outcomes, How did they get that way, and the key role of values (\S\ref{sec:7}).

\section{Hierarchies of structure and causation}
\label{sec:2}
All truly complex systems, including digital computers and life, are based in modular hierarchical structures 
\cite{Ellis_2016,Ellis_di_Sia} for good functional and evolutionary reasons \cite{Booch,Simon_2019}.

\paragraph{Orthogonal physical and logical hierarchies} Digital computers involve orthogonal physical and logical hierarchical structures (\cite{Ellis_2016}:\S2),  together forming \textit{computational artefacts} (\cite{Dasgupta_2016}:\S2) with the first supporting the second. Essentially the same is true in the case of human beings, because the brain is the basis of intelligence and understanding \cite{Mitchell_2023,Donald}.

\subsection{The physical hierarchies for digital computers  and for life}
The different levels of the emergent hierarchy of physical structure and function in digital computers \cite{Tannen,Ellis_2016,Dasgupta_2016} are shown on left hand side of \textbf{Table 2.1} and those for biology \cite{Ball_2023,Ellis_2016,Noble_2012,Campbell} on right hand side.  Different kinds of causation emerge at each level at each of these hierarchies, described by different kinds of dynamics and associated variables \cite{Anderson}. 

 \vspace{0.1in}
\begin{tabular}{|c|c|c|c|}
	\hline \hline
	& Digital computers & Human Life \\	
	\hline L10 & Internet &  Society \\ 
\hline 
	L9	& Local Network &  The Organism  as a whole \\ 
\hline L8 & Computer &  Physiological structures \\ 
\hline 
	L7	& Components &  Tissues  \\ 
	\hline 
	L6	& Gates & Cells   \\ 
	\hline 
	L5	& Transistors & Macromolecules, DNA \\ 
	\hline 
	L4 & Crystalline structures 	& Molecules
	\\ 
	\hline 
	L3	& Atoms & Atoms  \\ 
	\hline 
	L2	&  Ions, Carriers  & Electrons, Protons  \\ 
	\hline 
	L1	& Particle Physics & Particle Physics \\ 
		\hline \hline
\end{tabular} 
\vspace{0.1in}

\textbf{Table 2.1}:\label{Table2.1} \textit{The emergent hierarchy of physical structure and causation for digital computers (left) and human life (right)}  \cite{Ellis_2016}.\\

 The higher levels are characterised by larger scales of physical organisation;
furthermore processes come into play that cannot be described in lower level terms. Thus transistors are either conducting or not; cells are either dividing  or not; animals are either alive or not: and so on, Processes at level $\textbf{L+1}$ cannot be characterised in terms of variables defined at level $\textbf{L}$, which relate to different kinds of entities.
 
\paragraph{Artificial systems}  
 There are similar hierarchies in the case of artificial systems  such as watches, automobiles, aircraft, and cities \cite{Simon_2019,Arthur_2010}. 
 In the  case of digital computers,  the paper \cite{Ellis_Drossel_computers} carefully relates the higher level operations to the underlying quantum physics level. The core levels are \textbf{L5}: transistors \cite{transistors}, (\cite{Mellisinos}:1.9) and Logic Gates \textbf{L6} (\cite{Mellisinos}:1.9), where logical operations are derived from the underlying physics. 

\paragraph{Biology} 
On the life sciences side, quite different kinds of causation come into \textbf{}play at levels \textbf{L6} and upward \cite{Ball_2023,Campbell},  related to biochemical processes and molecular  biology \textbf{}(\textbf{L6}), cellular processes (\textbf{L7}), physiological processes (\textbf{L8}), mental and
psychological processes (\textbf{L9}) \cite{Mental_Causation,Deacon}, and social processes (\textbf{L10}) \cite{Berger}. Each of these levels is clearly causally effective; for example mental processes lead to the construction of aircraft and digital computers \cite{Ellis_2016}, which would not otherwise exist \cite{Ellis_2005}. The core level is the cellular level \textbf{L7} \cite{cells}, where all the functions of life first exist. Protein molecules act as computational elements in living cells \cite{Bray}.
Molecular processes underlie the brain  \cite{Ellis_Kopel}: membrane potentials alter ion channel shapes and so alter ion locations, and hence change the molecular Hamiltonian in a time dependent way. Messenger molecules alter protein conformations \cite{Lehn_2004,Lehn_2007} in response to higher level biological needs \cite{Hartwell_1999} conveyed by cell signaling networks \cite{Berridge}. The link to the underlying quantum physics is laid out in \cite{Karplus_1,Karplus_2014}.

\subsection{The reasons for modularity}
In order to obtain genuinely complex behaviour, a modular hierarchy of emergent levels \textbf{LN} is needed as in \textbf{Table 2.1}, with each level demonstrating different kinds of behaviour as we see in both digital computers \cite{Dasgupta_2016,Ellis_Drossel_computers} and life \cite{Ball_2023,Ellis_2016,Noble_2012}. Each level \textbf{LN} will be comprised of networks of modules at the next lower level \textbf{LN-1}. The reasons this is the necessary kind of structure to attain genuine complexity are discussed in \cite{Booch,Simon_2019}.

To generate genuine complexity associated with function, one must split a complex task up into simpler tasks, design units to
handle the simpler tasks, and then knit them together in networks so that  when
taken together they handle the complex task. The whole is done in a hierarchical way (sub-modules of modules do even simpler
tasks). At each level, 
 modules are characterised
by (a) Information Hiding, (b) Abstraction, (c) Interface Specification, (d) Modifiability, and (e) Multiple realizability.  This logic applies both  to physical and logical hierarchies.

\paragraph{Information Hiding} The module’s internal variables are hidden from outside view,
all that is required is that the module does what it is required to do, as viewed from
outside. The user does not mind what internal variables are chosen and what logic is
used.

\paragraph{Abstraction} An abstraction is an external description of what the module does in a
reliable way. This should include a name by which one can refer to the module, which
indicates what it does (it should have a unique identifier). 

\paragraph{Interface Specification} The interface between each module and the rest of the system
must be carefully specified, as in the case of the ISA: what control variables, parameters,
and data will be sent to the module, and what will be sent out to other modules
in the system from the module.

\paragraph{Modifiability} 
We can change the internal variables and even the internal logic of the module, as long
as the externally viewed functionality, as defined by the interface and abstraction, is
maintained. The system as a whole only cares that the module does what it is designed
to do, not how it does it (although some implementations are preferred over others if
they are either faster, or use less resources). This leads to multiple realisability in the hierarchy.

\paragraph{Multiple realizability}
A key feature of downward causation in both physical and abstract hierarchies is multiple realisation of higher levels at lower levels. Any higher level structure and dynamics arising out of lower levels  can be realised at the lower levels in numerous ways. Thus there is multiple realizability of every physical level \textbf{L} and its dynamics in terms of the lower levels \textbf{L-N} in \textbf{Table 2.1}. It follows that the higher level dynamics is not equivalent to lower level dynamics in any simple way.

In the case of digital computers this occurs for both the physical and logical hierarchies  \cite{Abramsky_multiple}. Higher level structures can be based on many different lower level structures in \textbf{Table 2.1}, e.g. using different computer chips with different instruction sets.
 Higher level programs can thereby be based on many different lower level languages (\textbf{Table 2.2}), e.g. using different operating systems. Even more than that: one can shift the boundary between hardware and software by different physical implementations \cite{Dasgupta_2016,Tannen}, so multiple realizability exists in this extended sense too. 

In the case of biology in general this is discussed in \cite{Farnsworth_et_al_multiple},
and its crucial role at the molecular level is powerfully presented in \cite{Wagner}. 
In the case of the brain, it is discussed in \cite{Figdor_Multiple,Ellis_Multiple}. Because of this multiple realisability with huge numbers of lower level elements (cells, molecules, atoms, particles) involved, the function and agency so patent at higher levels \cite{Ball_2023,Mitchell_2023} is indiscernible at these lower levels. That does not mean it does not exist \cite{Noble_Noble_2023}.

\subsection{Open systems and finiteness}
Biological systems are necessarily open systems for thermodynamic reasons \cite{Open1,Open2}. They must breathe and obtain food and liquid from the outside world because living things need energy \cite{energy}, and discharge solid and liquid waste into it. Furthermore there is a key issue relating to entropy: organization is maintained by extracting 'order' from the environment (\cite{Life}:\S 6)

Life also receives information from the outside  (light, sounds, smells) that influence their behaviour. They are structured so to be able to respond adaptively to such incoming data, and in turn act on it in various ways, including sending  information into the outside world. 

Digital computers are open systems in similar ways. They need to obtain electrical power from the outside world in order to function, and they dissipate heat into it. This is related to the physical limits on computation \cite{Landauer2,Landauer1}, in particular  the fact that all real computers have a finite memory, so items have to be erased from working memory in order that the calculation can continue. But such erasure is irreversible, and costs energy: any logically irreversible transformation of classical information necessarily involves dissipation of at least $kT\,ln(2)$ of heat per lost bit, where $k$ is the Boltzmann constant and $T$ the temperature \cite{Erasure1,Erasure}. Many computer programming languages (e.g. Lisp) have integrated garbage collection \cite{garbage} which carries out the function of freeing up memory which is not being used - which will generate heat. 

Computers also receive information from the outside world via the internet, which in turn enables them to act on the outside world in multiple ways.

\subsection{The three kinds of downward causation, and causal closure }

Downward causation occurs by three basic mechanisms \cite{Ellis_2023a}: (a) via time dependent constraints that shape lower level dynamics, (b) by creating, altering, or deleting lower level elements, (c) by coupling to the environment.

\paragraph{Non-unitary dynamics}
will occur when there is a time dependent potential term $V(r,t):\partial V(r,t)/\partial t \neq 0$ in the Hamiltonian, representing time-dependent constraints \cite{Ellis_Kopel}. The standard  uniqueness theorems \cite{Arnold,Arnold1} for Hamiltonian systems with time-independent potentials then don't hold: the initial state  does not uniquely fix the future evolution. This  happens in engineering  and biological systems, where higher level dynamics alters the context of lower level functioning.
Examples are changing voltages $V(t)$ controliing currents in MOSFET transistors in digital computers as a high level program is executed, the multi-scale dynamics of the heart involving cardiac sodium channel effects on action potentials and mechano-electric feedback in which the contraction
of the heart influences its electrical
properties \cite{Noble_2002}, and action-potential spike chains \cite{Fletcher} generated by voltage gated ion-channels \cite{Ellis_Kopel}. 

\paragraph{Creation, alteration,  and annihilation of elements} The laws of molecular biology or physics  determine specific outcomes in a given context - provided we know what entities they are operating on. In molecular biology, gene regulatory networks \cite{Karlebach_genes} operating in a specific context determine what proteins are present in a cell, which in turn determines what kinds of cells are present at specific locations in a body through developmental processes \cite{Carroll}.   The lower level elements change all the time in response to higher level dynamics \cite{Ellis_2023a}.

In solid state physics contexts, quasi-particles such as phonons are created by collective oscillations of a crystal structure \cite{Simon,QFT}. Ultimately they are created or destroyed by creation and annihilation operators that are an essential feature of Quantum Field Theory \cite{QFT,Kittel}. The key point then is when do these operators function? What turns them on and off? I return to this in \S5.

\paragraph{Coupling to the environment}
As well as imposing time-dependent constraints, which set the context in which the time evolution of a system takes place, the environment influences a system also via ongoing exchanges of energy, information, and possibly matter \cite{Ellis_closure,Ellis_Open}. No system is fully isolated from these influences. This is a further key form of downward causation, whereby arrows of time get injected into local systems via contact with heat baths which in turn are in contact with the external environment up to the Universe \cite{CWC,Ellis_Drossel_time}.\\

\paragraph{Computers and users} A key point here is that the two themes of this paper - the way computers function and the way intelligent life functions - are linked by the fact that the computer user is part of the environment of the computer, who feeds information in that then determine its operation \cite{Dasgupta_2016}. This is illustrated in  \textbf{Table 2.2}. It is how mental causation (decisions made by the user) controls the operation of computer hardware, influenced by the society around \cite{Berger}, right down to the quantum physics level.

\begin{table}[h]
\centering 
\begin{tabular}{|c|c|c|}
	\hline \hline
	& Society &  \\	
	\hline 
  &  $\Uparrow$  \hspace{0.3in}  $\Downarrow$ & \\ \hline
	& Computer User &  \\	
	\hline 
  &  $\Uparrow$  \hspace{0.3in}  $\Downarrow$ & \\ 
\hline 
& Application software, data &  \\ 
\hline 
& $\Uparrow$	 \hspace{0.3in}  $\Downarrow$ & \\ 
	\hline 
&  Operating system  & \\ 
	\hline  
& $\Uparrow$	 \hspace{0.3in} $\Downarrow$ & \\ 
	\hline 
& Computer Hardware 	&  \\ 
	\hline 
 \hline
\end{tabular} 
\end{table}

\textbf{Table 2.2} \label{Table 2.2}\textbf{Relation of user to computer} \textit{The computer is an open system, with its operation controlled by the user in the context of society}.\\

\noindent  Knowledge of the internal state of the computer at any time $t_1$ will not determine its state at a later time $t_2 > t_1$ because of this openness. 

\paragraph{Causal closure}  Because of the combination of these upwards and downwards forms of causation, it is not true that any underlying physical level is by itself causally closed: all the interacting levels contribute to the final result and affect the outcome, because of this causal chaining: it would not happen if they were not all there. That is why every level is causally effective \cite{Noble_2012}. Actually one needs only go down to the level of electrons and protons, which is the lowest effective physical level, because of the existence of quantum protectorates \cite{Laughlin_Pines}: the lower levels (quarks, gluon, and so on) are irrelevant to outcomes in digital computers and life.

\subsection{The Logical Hierarchy for digital computers}

In the case of digital computers, there are two modular hierarchical structures: the physical hierarchy discussed above (\textbf{Table 2.1}), and a logical hierarchy  supported and enabled by the structure and function of the physical hierarchy (\cite{Ellis_2016}:\S2, \cite{Ellis_Drossel_computers}). They can be regarded as orthogonal to each other, because although the logical hierarchy is realised by the physical hierarchy, the content of the logical hierarchy is determined by the software and data loaded, not by the nature of the physical hierarchy.

 \vspace{0.1 in}
\begin{tabular}{|l|l|l|l|}
	\hline \hline
Level	& Language and Logic & Downward Process & Agent \\	
	\hline 
 $M_6$ & $L_6$: Applications Program &  $\Downarrow$ Logic and data & Program Code \\ 
\hline 
$M_5$	& $L_5$: High level Language & $\Downarrow$ Translation: Compiler & Syntax, semantics \\ 
\hline 
$M_4$	& $L_4$: Assembly Language &  $\Downarrow$ Translation: Assembler & Syntax, Semantics \\ 
	\hline 
$M_3$	& $L_3$: Operating system level & $\Downarrow$   Partial Interpretation & OPerating system \\ 
	\hline  
$M_2$	& $L_2$:	Machine Language & $\Downarrow$ Interpretation & ISA \\ 
	\hline 
$M_1$ & $L_1$: Microprograms 	& $\Downarrow$ Directly executed & Instruction \\ 
	\hline 
$M_0$	& Digital Logic Level (binary)  & \,Hardware & Gates  \\ 
		\hline \hline
\end{tabular} 
\vspace{0.1in}

\textbf{Table 2.3}:\label{Table2.3} \textit{Logical hierarchy for a digital computer}  \cite{Tannen,Ellis_Drossel_computers}.\\

he logical hierarchy (\textbf{Table 2.3}) represents the tower of virtual machines $M_I$ \cite{Tannen} each with an associated language $L_I$. Multiple realizability occurs between every pair of logical levels because different languages can be chosen at lower levels, expressing exactly the same logic (the algorithm) as do the higher levels, but in a different language. Higher level data also gets re-expressed in the relevant lower level notation. The downward arrows are implemented by compilers or interpreters, according to context,

Here for example $L_6$ might be a word processor, internet browser, image processor, AI program, and so on; $L_5$ might a language such as FORTRAN, $C^{++}$, JAVA, PYTHON, and so on. This logical hierarchy is implemented by the physical hierarchy (\textbf{Table 2.1}), but not in a 1-1 way. In a particular context of computer use, every level of the physical hierarchy is at work representing every level of the logical hierarchy. \\

A crucial point is that the data and logic represented by the logical hierarchy is not in any way restricted by the 
physical hierarchy: the physical implementation does not in any restrict the logic it represents, provided it does not use up too much memory or take too much time to process.
\begin{quote}
    \textit{\textbf{The principle of arbitrary content: digital representations}} \textit{The logical hierarchy of a digital computer can represent arbitrary logic and data, which is re-expressed in different notations at each of the logical levels, until it is expressed in binary notation (0s and 1s) at the lowest level \textbf{L0}. This flexibility is based in the possibility of using arbitrary  software at the higher levels to control flow of electrons in gates.} 
\end{quote}
 That is why these hierarchies are called ``orthogonal''. This is how abstract causation works in this context: abstract elements (programs, data) act downwards to control events at the micro-physical level (\cite{Ellis_2016}:\S2, \cite{Dasgupta_2016}), indeed right down to the quantum level \cite{Ellis_Drossel_computers}).  \\

This underlies the extraordinary usefulness of digital computers in the world: any data whatever,
representing pictures, diagrams, videos, scientific measurements, documents, cooking recipes, engineering variables can be represented digitally and manipulated logically to calculate probabilities and possibilities, and even to determine physical outcomes such as manufacturing processes in automated factories, generating 3-d printed objects of any kind, and controlling systems such as chemical plant or an aircraft or automobile. 

\subsection{The Symbolic Brain and Abstract thought}
A key feature separating human beings from the great apes and all other life is that we are a symbolic species \cite{Deacon} able to represent and analyse the world in symbolic terms. This occurs at level \textbf{L9} in \textbf{Table 2.1}. It is enabled by the physical structure of our brain: an enormously complex neural network \cite{Kandel} with the capacity to learn, analyse, predict, and make choices. Any book, paper, article, or speech in any language is an abstract modular hierarchical structure represented in physical form, as are the categories of objects and actions they express, as set out in dictionaries and encyclopedias. This logical content is in each case representable in many different ways both at each level, and at lower levels: the basic principle of multiple realisability holds again in this context, whereby for example a noun phrase can act as a noun. 
This extremely flexible  structure has evolved for good functional \cite{Carruthers} and evolutionary \cite{Mitchell_2023,Sterelny,Miyagawa} reasons. It allows us to act on the world with agency \cite{Mitchell_2023}, for example creating entities such as digital computers, which would not otherwise exist.  \\

\noindent In parallel with the case of computers just discussed, the data and logic represented by the functioning of the brain is not restricted by the its physical structure, apart from memory limitations and processing time.
\begin{quote}
    \textit{\textbf{The principle of arbitrary content: brain plasticity}} \textit{The physical brain can represent, analyse, and make predictions from arbitrary logic and data: any thoughts about any topic can occur. These operations of the mind result in, and are based in, brain plasticity \cite{plasticity}: modifications of neural network connection weights in the brain enable them, which do not in any way restrict what is represented.}
\end{quote}
This flexibility  is what underlies the enormous causal power of abstract thought and associated mental causation \cite{Mental_Causation}, which has transformed the world around us, creating societies, agriculture, money, economic systems, aircraft, factories, buildings, water supplies, and so on, as well as transistors, integrated circuits, digital computers, and the internet.

\paragraph{\textbf{A challenge}} We do not know what the equivalent of the logical hierarchy \textbf{Table 2.2} is in the case of a brain. Perhaps it includes some manner of representation of thoughts via coding in adaptive resonant circuits \cite{Grossberg,Grossberg1}, chaining down to  the structure (logic?) of action potential spike chains \cite{Fletcher} and then to physical implementation via voltage gated ion channels opening and closing \cite{Ellis_Kopel}, based in the underlying quantum chemistry \cite{Karplus_2014}. However to clarify this also requires engagement with the cognitive neuroscience of language \cite{Kemmerer}. 

\subsection{Branching dynamics at every level}
A key feature of both the physical and logical hierarchies is that branching dynamics occurs at every level, in an integrated way between levels. 

\paragraph{Logical branching: Computers}   In the case of a digital computer, a high level program $L_6$ will have a statement of the logical form (\cite{Dasgupta_2016}:36)
\begin{equation}\label{eq:If_then}
    IF \,\,\, \{\textbf{T1}\}\,\,\,THEN\,\, DO \,\,\, \{\textbf{A1}\} \,\,\, ELSE\,\,\,  DO \,\,\, \{\textbf{A2}\}
\end{equation}
where \textbf{T1} is a logical truth statement (e.g. $\textbf{T1} := X \geq Y$), while \textbf{A1} and \textbf{A2} are two alternative actions. This can represent an arbitrary real world issue: for example $X$ might be money owing,  $Y$ a credit limit, \textbf{A1} is ``refuse loan'' and \textbf{A2}  ``grant loan''; or  $X$ might be cost of holiday in Rio,  $Y$ amount saved for a holiday, \textbf{A1} is ``stay home'' and \textbf{A2}  ``go to Rio''. As pointed out in \S2.5, any arbitrary issue whatever can be represented in this way.

\paragraph{Physical branching: Computers}
Corresponding to the logical branching, physical branching takes place at each level
in \textbf{Table 2.1}:
\begin{itemize}
    \item At Transistor Level the branching is ON or OFF, depending on the gate voltage $V(t)$.
\item At Gate Level the branching is in terms of truth values of basic logical operations:
AND, NOT, NOR which can then lead to branching in comparators, adders, decoders,
etc. (\cite{Mellisinos}: pp.37-58;\cite{Tannen}: pp. 135-164), with outcomes depending on the data. 
\item At Computer Level branching occurs via CPU control of the basic instruction FETCH-EXECUTE
cycle (\cite{Mellisinos}:p.75, \cite{Tannen}:pp.173–202), with activation of different instruction
memory and data memory locations and output modules via a bus (\cite{Tannen}:pp.176–220)
controlled by a clock. Branching of electron flows in the data bus occurs according to
the relevant data.
\item At Network Level branching is  via requests represented by a Hypertext
Transfer Protocol (HTTP), sent between computers identified by their Internet Protocol
address (IP address), as for example sending a request to Google for information. 
\end{itemize}

\paragraph{The relation between them}
Logical branching governs physical branching, which is demonstrable by changing
the program loaded. Outcomes alter, although the physical structure involved (the
computer hardware) is identical whatever the logic involved. This happens via compilers (\cite{Dasgupta_2016}:47,\cite{compiler}) and interpreters \cite{Tannen}
that chain logic down to the machine code level.

The logic of the algorithm \cite{Knuth}, e.g. Bubblesort, is preserved during the downward chaining process, as
it gets rewritten in different languages \textbf{LI} with different variables and syntax (\textbf{Table 2.3}).
The abstract becomes physical at the machine level, where digital logic is expressed in a timed sequence of electron flows that turn transistors ON or OFF \cite{Tannen,
Mellisinos}.
The bits are translated into electric voltages controlling  the transistor logic \cite{Ellis_Drossel_computers,Mellisinos}: 
\begin{equation}\label{eq:threshold}
    IF \,\,\,\{ V(t) > V_{threshold}\} \,\,\,THEN\,\, \{\textbf{current flows}\}   \,\,\, ELSE\,\,\,  \{\textbf{not}\}.
\end{equation}
This is implemented physically as explained in \textbf{Table 4.1} in \S4.1. 

\paragraph{Logical and physical branching: the Brain} Essentially the same applies in the brain. The branching logic (\ref{eq:If_then}) may  apply at the mental level with regard to some option we  consider,
or the simpler case
    \begin{equation}\label{eq:If_then1}
    IF \,\,\, \textbf{T1}\,\,\,THEN\,\, DO \,\,\, \textbf{A1}
\end{equation} 
may apply, where for example \textbf{T1} is ``I want to learn General Relativity'' and \textbf{A1} is 
``Learn tensor calculus first'' (with the implict logic `If NOT \textbf{A1} then NOT \textbf{T1}''). This will chain down in ways we do not understand to specific action potential spike chains in neuronal axons, enabled at the molecular level by the logic ` 
\begin{equation}\label{eq:threshold1}
    IF \,\,\, \{V(t) > V_{threshold}\} \,\,\,THEN\,\, \{allow \,\,ion\,\, flow\} \,\,\,  \,\,\, ELSE\,\,\,  \{not\}
\end{equation}
where $V(t)$ is the voltage difference across the axon membrane and voltage gated ion channels open or close to implement this logic \cite{Ellis_Kopel}. At an even lower level this will be implemented by quantum chemistry interactions that control molecular shape \cite{Karplus_2014}. 

\paragraph{Downward logical and physical branching}
In both cases, logical branching at the highest level (algorithms, thoughts) drive logical branching consistently at all lower levels by the mechanisms discussed further in Sections 4 and 5, with multiple realizability occurring at each downward step,  until either contextual wavefunction collapse or particle creation and annihilation operations shape the outcomes at the quantum level. 

\section{Quantum linearity, local wave functions, and and wave function collapse}\label{sec:3}
Both digital computers and life are enabled at the bottom physical levels by quantum physics as characterised by the Schr\"{o}dinger equation. However this is a linear equation, so it is something of a puzzle as to how this allows   complex structures such as digital computers, cats, human beings, and societies to come into being \cite{Ellis_2023}. None of them is remotely linear: their dynamics is not unitary.

\subsection{The Schr\"{o}dinger equation}
The Schrödinger equation \cite{Landshoff_et_al,Laughlin_Pines,Isham_2001} is 
\begin{equation}\label{eq:Schroedinger}
 i \hbar {\frac{\partial}{\partial t}} |\psi(t)\rangle   = \hat{H}|\psi(t)\rangle
\end{equation}    
where $\hat{H}$ is the Hamiltonian operator. 
In position space representation for a many particle system this is \cite{Drossel_Limits}
\begin{equation}\label{eq:Schroedinger1}
 i \hbar {\frac{\partial}{\partial t}} |\psi(\vec{r}_1,...,\vec{r}_N,t)\rangle   =\left[-\frac{\hbar^2}{2m}{\frac{\partial^2}{\partial x^2}}+V(\vec{r}_1,...,\vec{r}_N,t)\right]|\psi(t)|\psi(\vec{r}_1,...,\vec{r}_N,t)\rangle
\end{equation}        
This is linear because there is no source term on the right in addition to the $\hat{H}$ term, and  $V(\vec{r}_1,...,\vec{r}_N,t)$ is independent of $|\psi(\vec{r}_1,...,\vec{r}_N,t)\rangle$.   

\subsection{Linearity and local wave functions}

Because of this linearity, quantum theory for a single wave function $|\psi(t)\rangle$ cannot describe non-linear systems \cite{Ellis_2012}. However, these exist in the physical world, both in the context of  digital computers and in biology at each emergent level $\textbf{L} \geq \textbf{L4}$ in the hierarchy of structure (\textbf{Table 2.1}). Consequently, as I now argue, quantum physics can only hold locally \cite{Drossel_Limits,Drossel_2017,Ellis_2023} 
via local wavefunctions that hold only in restricted domains. 

\paragraph{Non-linearities} Where do essential non-linearities occur in digital computers and in biology? - in five essential cases:
\begin{itemize}
\item \textit{Hierarchies}, as discussed in detail above. An intricate non-linear relation of functions exists between all the levels in \textbf{Table 2.1} and \textbf{Table 2.3}.
    \item \textit{Feedback control loops}/cybernetic systems such as aircraft autopilots, and homeostatic systems that are crucial to biology at micro and macro levels
    \item \textit{Networks}: these occurs in digital computer integrated circuits; in biology, in gene regulatory networks, metabolic networks, and neural networks. 
    \item \textit{Heatbaths} which have a coherence timescale and physical length \cite{Drossel_2014,CWC}
     \item \textit{Creation and deletion of lower level elements} In digital computers, creation of quasi-particles that shape conduction properties of transistors; in biology, reading of genes to to produce proteins
\end{itemize}
None of this is unitary: but they certainly happen. So how do all the non-unitary dynamics characterised in \textbf{Table 2.1} arise? 
The key question is, \textit{What are the limits of quantum mechanics?} \cite{Drossel_Limits}, The answer has to be that while quantum physics locally underlies all emergent physics and chemistry everywhere, it does not do so by wave functions with an unrestricted domain of applicability, such as a single global wave function for the universe. In fact only local wave functions exist, each valid only in a restricted domain \cite{Ellis_2023}. The proposal is based in the way that in General Relativity Theory,  there are in general no global coordinates systems covering a spacetime manifold; rather an atlas of local coordinates systems exist that taken together cover a maximally extended spacetime \cite{HE}. The same logic applies in quantum physics.

\paragraph{Local wave functions}  
The idea is that only local wave functions $| \psi \rangle_\alpha$ exist that are each valid only in local 
domains ${\cal U}_\alpha$ where the dynamics expressed by $| \psi \rangle_\alpha$ is linear \cite{Ellis_2012,Ellis_2023}. The notation 
\begin{equation}
    | \psi \rangle_\alpha = | \psi \rangle_\alpha ( {\cal U}_\alpha) 
\end{equation}
expresses this limited domain of validity.  Any non-linear dynamics such as in the cases just mentioned will be split up into linear parts each described by such a local wave function $| \psi \rangle_\alpha$; the whole system will be covered by the union of associated wave function domains ${\cup}_\alpha( {\cal U}_\alpha)$. In the overlap ${\cal U}_{\alpha\beta} := \{{\cal U}_\alpha \cup  {\cal U}_\beta \}$ between any two of them, they will be related by an invertible  transformation
\begin{equation}    
| \psi (x,t) \rangle_\alpha = 
f_{\alpha\beta} (
| \psi (x,t) \rangle_\beta)
\end{equation}
and associated Bogoliubov transformation between the relevant Hilbert spaces. Note that ${\cal U}_\alpha$ may be time dependent: ${\cal U}_\alpha = {\cal U}_\alpha(t)$. 

\subsection{Implications}
Consequently there is no single wave function for a living cell in a human body, a cat, a brain, or a person, for non-linear processes occur in all of them. 

\paragraph{Schr\"{o}dingers' Cat} 
In the context of this famous thought experiment \cite{Leggett}, the ``equation''
\begin{equation}
    | \psi \rangle_{cat} = \alpha | \psi \rangle_{alive} +  \beta | \psi \rangle_{dead}
\end{equation}
is nonsensical, as none of those wave functions exist: the cat's physiology is highly non-linear. In the typical Schr\"{o}dinger cat experiment, the only quantum entity described by a single wavefunction  is the excited atom. All the rest - the cat, the cage, the poison vial, the hammer, the particle detector are classical. The cat is never in a superposition of being both alive and dead.

\paragraph{The Wave Function of the Universe and the Everett interpretation} Because the Universe contains living cells, cats, and human beings, for each of which no single wave function exists, there is also no single wave function  $| \psi \rangle_{Universe}$ for the Universe as a whole considered at all scales, as is routinely claimed. This causes problems for the Everett interpretation of quantum physics which envisions $| \psi \rangle_{Universe}$ splitting all the time into multiple branches \cite{Isham_2001,Many_Minds1}. This proposal relies on existence of  a single linearly evolving wave function for everything, taking for granted that it exists. This is highly implausible, and at a minimum requires serious justification. 
Similar problems arise for any ``many minds'' interpretation  \cite{Isham_2001,Many_Minds1,Many_Minds}, whether related to the Everett interpretation or not. They assume existence of a single wave function $| \psi \rangle_{brain}$ for each brain, which on the basis of the above arguments I claim does not exist.

\paragraph{An effective Copenhagen Interpretation}
Nor is there a single wave function for the apparatus involved in any quantum physics experiment. Consider for example Alain Aspect’s paper on the Bell test (\cite{Aspect}: Figure 1). All is classical: the source, beam splitters, detectors, coincidence detectors), except a few interacting particles ($\nu_1$ and $\nu_1$) described by wave functions. All the rest are conceived of and described in classical terms. To describe them in
quantum terms would make the analysis impossible: one would have to model each of the elements of the apparatus quantum mechanically, and they would 
each have no definite state.
This confirms the claims in \cite{CWC} that wave function collapse (the quantum to classical transition that is necessarily involved in any physics experiment which produces a classical outcome which can be analysed) is 
enabled by the classical context, as discussed in section \S3.4. 

\subsection{Collapse of the wave function}\label{sec:collapse}

 A key aspect of downward causation  is contextual wavefunction collapse  \cite{CWC}.  

Quantum theory has two parts (\cite{Penrose_Road}:527-533): unitary wavefunction evolution $U$, plus wave function reduction $R$. The latter leads to definite physical outcomes, and so is associated with the passage of time. 
 As the very purpose of the wave function is to determine probabilities of classical outcomes, quantum theory  means nothing physical unless wave function reduction $R$ occurs \cite{Penrose_Road,Ellis_2012}. However it is often omitted from discussions of quantum physics.

A simplified description of the process $R$ is as follows: When an event $R$ happens, a wave function $|\Psi\rangle$ that is a superposition of orthonormal  eigenstates  $|u_n\rangle$ of some operator is projected to a specific eigenstate $N$ of that operator:
 \begin{equation}\label{eq:collapse}
R:\,\,|\Psi\rangle(t_0) = \Sigma_n c_n |u_n\rangle \,\rightarrow |\Psi\rangle(t_1) = \alpha_N |u_N\rangle. 
\end{equation}
where $\alpha_N$ is the eigenvalue for that eigenvector. In reality, this process is far more complex, as discussed in \cite{CWC} (see below). 
There is irreducible uncertainty in this irreversible process:  the specific outcome $N$ that occurs is not determined uniquely by the initial state $|\Psi\rangle(t_0)$ \cite{Ghirardi}. However the statistics of the outcomes is reliably determined by the Bohr rule:  the probability $p_N$ of the specific outcome $|u_N\rangle$ occurring is given by  
\begin{equation}\label{eq:Born}
 p_N = |c_N|^2.
 \end{equation}
The projection process (\ref{eq:collapse}) occurs in laboratory experiments such as the 2-slit experiment, and so is often taken as being related only to such experiments. However it also occurs all the time when physical interactions take place,  such as nucleosynthesis in the early universe (when no observers were around), when a photon is registered by a CCD, and when a photon hits a chlorophyll molecule in a leaf and hence releases an electron that starts a cascade of biochemical reactions during photosynthesis in plants. 
Thus in many real-world contexts non-unitary events (\ref{eq:collapse}) take place that cannot be described by (\ref{eq:Schroedinger}) \cite{Isham_2001,Penrose_Road}, and are not associated with a laboratory experiment. \\

\textit{Contextual Wavefunction Collapse} (CWC) \cite{CWC} proposes that a nonunitary, stochastic  process (\ref{eq:collapse}) takes place and obeys the Born rule (\ref{eq:Born}), with the way this happens being determined by the  local physical context. Indeed this is obviously the case:  specific apparatus may measure energy or polarisation, and outcomes depend on this choice. In the latter case the  direction of polarization measured can be chosen at will, again altering outcomes \cite{Susskind}. 

I will not repeat in detail the proposal \cite{CWC} here, but summarise as follows.  For concreteness that paper  focuses on the case of a photodiode, detailing the processes whereby 
\begin{itemize}
    \item A photon is absorbed and an electron lifted from a bound state to the conduction band; 
    \item An internal heat bath made of phonons interacts with the electron in the conduction band and localizes it;
    \item Interaction with an external heat sink makes the process irreversible and introduces an arrow of time;
    \item An electrical field moves the electron from the site of the excitation, with interactions with other electrons triggering a macroscopic avalanche;
    \item Detection of the resultant macroscopic current by classical processes results in movement of a pointer or lighting of a bulb;
    \item A reset step involves an external heat sink and requires energy; this makes the process irreversible. The associated direction of time ultimately arises from the expansion of the universe.  
\end{itemize}
In effect this is a specific form of the Copenhagen interpretation \cite{Isham_2001} where the macro apparatus is classical rather than quantum. The reason for this is limitations of the domain of validity of  wave functions $|\Psi\rangle$ \cite{Ellis_2012}, as just discussed, and in particular the fact that a heat bath cannot be described by a many particle wave function \cite{Drossel_2017}. Any real macro apparatus involves heat baths and so is a classical entity, even though it emerges from a structure that has quantum properties on the microscopic scale.  

An alternative that in many ways has a similar motivation is \textit{Relational Quantum Physics}  \cite{Rovelli}. This is an innovative and interesting proposal,  but the Contextual Wavefunction Collapse proposal is tied in more closely to real physical contexts, such as the operation of photo-diodes or transistors. It would be interesting to see Relational Quantum Physics  applied in specific contexts.

\section{Quantum Theory and Time dependent constraints}\label{sec:4}
The first form of downward causation mentioned above (\S2.4) is the existence of higher level time-dependent constraints that shape lower levels outcomes. 

The operational question is, Where does $V(\vec{r}_1,...,\vec{r}_N,t)$ in (\ref{eq:Schroedinger1}) come from? This potential represents downward causation via constraints representing higher level contexts \cite{Noble_2012,Juarrero}. When quantum tunneling is involved it may be characterised classically as a potential well. 
In the case of digital computers and the quantum chemistry underlying life, it occurs via the Born-Oppenheimer approximation \cite{BO}  suitable to the context (crystals and molecules respectively) and their time dependent emergent dynamics.

 \subsection{Digital computers and transistors}\label{sec:transistors}
 At a high level time dependent constraints represent voltages resulting from currents resulting from abstract computation that determine when transistors are on or off \cite{Dasgupta_2016}. The way this works out in the case of digital computers at the lower levels is set out in (\cite{Ellis_Drossel_computers}:\S 4.1, Table 3), reproduced here as \textbf{Table 4.1}. 

\vspace{0.1in}
\begin{tabular}{|c|c|c|c|}
	\hline \hline
	& \textbf{Level} & \textbf{Structure} & \textbf{Outcomes} \\	
	\hline 
 \textbf{T5} & Gates       &  Transistor combinations & Boolean Logic \\ \hline
 \textbf{T4}	& Transistors & Base, emitter, collector & ON/OFF \\  
        &      & Carrier channels & \\ \hline
 \textbf{T3} & Crystal structure &  
 Ions: symmetry breaking  &  Phonons, band structures \\ 
   &   & by impurities & \\ \hline
 \textbf{T2} & Electron population &  Densities, average flows,  & Electron diffusion, current \\
  & & resistance & \\ \hline
\textbf{T1}	&  Individual ions, electrons & Ion bonding; electron velocities,  & Electron drift  \\
& & collisions & \\
		\hline \hline
\end{tabular} 
\vspace{0.1in}

\textbf{Table 4.1:} \textbf{The lower physical levels in a digital computer} \cite{Ellis_Drossel_computers}.\\

\noindent The levels are as follows.

\textbf{Level T5}: \textbf{Gates} Transistors are linked via wires and resistors to form the basic logical gates,  perhaps combined to form a single integrated set of transistors.

\textbf{Level T4}: \textbf{Transistors}  are based in semiconductors such as silicon, doped with donor (n) or acceptor (p) impurities.  An applied voltage $V(t)$ on the gate attracts electrons from the source and thereby opens a conducting channel between the source and drain, through alterations to the chemical potential and depletion region. 

\textbf{Level T3}: \textbf{Crystal Structure, Phonons, Electronic Bands}  The crystal structure with
its  particular symmetries \cite{Kittel} and degrees of freedom gives rise to
phonons and the electronic band structure.  Constraints representing the periodicity of the crystal lattice  underlie the use of Bloch's Theorem (\cite{Kittel}:p.179) 
Depending on the distance between the
bands and their filling, one obtains the distinction between conductors, insulators, and
semi-conductors.

\textbf{Level T2}: \textbf{Electron Population} Electron/carrier flow is due to diffusion: 
density gradients leading to depletion regions, and to  drift due to an electric potential $V(t)$,
leading to a current. Resistance occurs due
to collisions with impurities, phonons, and other electrons. Electron conduction is
time asymmetric because of interaction with a heat bath. Due to its dissipative nature, this is not a Hamiltonian process.
an emergent description (a Boltzmann equation) is used: a phenomenological theory for this level.  

\textbf{Level T1}:\textbf{Individual ions and electrons}  The description at this level is based
on a Hamiltonian for the ions and electrons. 
The electrons are separated into conduction band electrons (essentially unbound and so free to move) and valence band electrons (closely bound to ions and so localised).
The Born–Oppenheimer (adiabatic) approximation \cite{BO} is used in the quantum Hamiltonian with a symmetry-broken ground state: the crystal lattice structure which leads to phonons existing \cite{Kittel}. The electron equation is used to obtain the electronic band structure.  
Electron-lattice interactions occur via phonons (\cite{Kittel}:\S2). To model electron-phonon interactions explicitly, a quantum field theoretical formalism is required based on creation and annihilation operators for electrons and phonons.  

\textbf{Interlevel effects} The steady state situation when the transistor is   either conducting or not depends on whether  the applied voltage is above a threshold. The  electric field effects are modelled by adding the gate voltage $V(t)$. This leads to a potential
energy term $H_V(t)$ in the Hamiltonian of the electrons:
\begin{equation}
  H_V(t) := \sum_i e V(\textbf{r}_i(t))
 \end{equation}
where the Level \textbf{T4} variable $V(t)$ determines the Level \textbf{T1} variables $V(\textbf{r}_i(t))$ in a
downward way. This leads to a displacement of the electrons until a new equilibrium is
reached where the electrical field created by the modified charge distribution cancels
the electrical field due to the gate voltage. 
This alters the band structure 
and thereby either creates
a conduction channel by changing the depletion zone, or not, according to the bias
voltage applied.

\paragraph{\textbf{The outcome}} Outcomes are determined by the time dependent function $V(t)$ determined by the
machine code (Level $M_1$ in Table 2.3), and applied in the context of the specific detailed
structure of the transistor concerned. This is all driven by the programs and data loaded as indicated in \textbf{Table 2.3}. 

This is how the logic represented in the abstract structure of the computer program, expressed in machine code \cite{Dasgupta_2016} controls the underlying physics at the electron level; the electron dynamics is no longer unitary. Quantum level outcomes are shaped by abstract algorithms and data
\cite{Ellis_Drossel_computers}. Different programs or data result in different physical outcomes.

\subsection{Quantum chemistry and Molecular shape}
 In living systems, at the molecular level change of shape of molecules in cell signalling pathways (\cite{Berridge}:\S 2) governs the dynamics \cite{Lehn_2004,Lehn_2007}. However the relevant molecular processes \cite{Ball_2023} implemented by metabolic networks and gene regulatory networks  are driven downwards by physiological needs \cite{Noble_2012} and developmental processes \cite{Carroll}. The underlying quantum chemistry potential  $V(\vec{r}_1,...,\vec{r}_N,t)$ in (\ref{eq:Schroedinger1}) is consequently time dependent, and hence allows logical branching  to arise from the underlying physics \cite{Ellis_Kopel}.

Changes in molecular shape are the key active factor in biological function at the molecular level \cite{Ball_2023,Lehn_2004,Lehn_2007}. However shape is a classical variable, not a quantum one. 
A central issue  is, how does the classical concept of shape, key to occurrence of
these processes, arise out of quantum theory? This is  discussed in depth in \cite{Ramsey_1997} and (\cite{Bishop_Ellis}:§5.2). In agreement with the theme of this paper, one can claim that molecular shape is shaped by the
environment \cite{Amann_1991,Amann_1993}.

The way that change of shape is calculated in terms of quantum theory is presented by Martin Karplus in \cite{Karplus_2014}, again emphasizing (like \cite{CWC}) that quantum calculations in complex situations rely on classical concepts to a large degree.
The number of atoms in a macromolecule \cite{Lehn_2007} is huge. To handle their
complexity requires multiscale models. Karplus states  \cite{Karplus_2014},
\begin{quote}
  \textit{  “To develop methods to study complex chemical systems, including biomolecules, we
have to consider the two elements that govern their behavior: 1) The potential surface
on which the atoms move; and 2) the laws of motion that determine the dynamics of the
atoms on the potential surfaces. ...although the laws governing the motions of atoms
are quantum mechanical, the key realization that made possible the simulation of the
dynamics of complex systems, including biomolecules, was that a classical mechanical
description of the atomic motions is adequate in most cases.”}
\end{quote}
Thus one does not actually solve (\ref{eq:Schroedinger1}) in this case. A classical background provides a basis for determining the quantum states, as is generally true when wavefunction collapse determines quantum outcomes\cite{CWC}. 

How does this relate to biological function? Karplus states further \cite{Karplus_2014},
\begin{quote}
   \textit{   “First, evolution determines the protein structure, which in many cases, though not
all, is made up of relatively rigid units that are connected by hinges. They allow the
units to move with respect to one another. Second, there is a signal, usually the binding
of a ligand, that changes the equilibrium between two structures with the rigid units in
different positions. ...This type of conformational change occurs in many enzymes as
an essential part of their mechanism. Thus one does not solve the Schr\"{o}dinger equation in this case. A classical background provides a basis for the quantum}
states.\end{quote}
This clearly parallels the situation for digital computers, even though in a quite different context.

\section{Quantum Field Theory: Particle Creation and Annihilation}\label{sec:5}
The second key form of downward causation mentioned above (§2.4) is the creation, modification, or deletion of lower level elements \cite{Ellis_2023a}. This is fundamental to both digital computers and biology.  

At the quantum level, this is based in the fact that  the creation and annihilation of particles is central to Quantum Field Theory \cite{QFT,Kittel}. These opeartions are implemented by creation and annihilation operators $\textbf{a}$, $\textbf{a}^\dagger$.  

\subsection{Digital computers}
Such creation and annihilation occurs in electron-phonon interactions, as for example occurs in transistors at \textbf{Level T1}, as just discussed. This is occurs via creation and annihilation operators \cite{Ellis_Drossel_computers,Transistor-Create}.

 \subsection{Biology}
Creation, alteration, and deletion of lower level elements
is a key feature of biology \cite{Ball_2023} for example in developmental processes \cite{Carroll} controlled by gene regulatory networks \cite{Karlebach_genes,GRN,GRN1,GRN2} that create proteins as needed by the higher level physiological context. 
At the quantum level this is occurs via creation and annihilation operators 
as in molecular structure calculations \cite{Molecule-Create,Molecule-Create1}.

It also happens in photosynthesis \cite{photosynthesis,rhodopsin} where photons are absorbed and electrons emitted.

\subsection{When does it happen?}

When do creation/annihilation operators act? Standard QFT texts emphasize existence of these operators, but do not make clear when they change outcomes \cite{QFT}. It seems likely that this timing is controlled downwardly by a classical context, as suggested by these examples. \\

\textbf{Hypothesis}: \textit{The action of creation and annihilation operators in QFT is controlled in a downward way by the physical context, as occurs generically in the cases of Contextual Wavefunction Collapse \cite{CWC}, and specifically in the quantum physics underlying digital computers \cite{Ellis_Drossel_computers} and macromolecular dynamics \cite{Karplus_2014}. }

\section{Open systems and the arrow of time}\label{sec:6}
The third form of downward causation mentioned above (§2.4) is ongoing interaction with the environment.   

Open systems need a heat sink, and this is provided by the dark night sky \cite{Penrose_Road,Penrose_Fashion} and
indeed by the low temperature of the sky in all directions but that of the Sun during the day. This enables heat baths to radiate low grade energy into the sky. The existence of a dark night sky is called Olber's Paradox, because in static universes the entire sky potentially would be at the same temperature as the surfaces of stars \cite{Bondi}. The solution to this paradox is that the universe is expanding and evolving \cite{Harrison,Ellis_Sciama}. Its thermal history results in the present cosmological background radiation temperature of $2.73K$ \cite{Dod03,Peter_Uzan}. But because of the greenhouse effect \cite{greenhouse}, the sky on Earth is at an effective temperature of about $15K$ - which is what life has adapted to. 

Heat baths cannot be described by a many particle wave function \cite{Drossel_2017} but are a crucial link between an open system (which all digital computers and living systems are) and the environment. They are the way that the thermodynamic arrow of time (determined by the cosmological direction of  time \cite{EBU,Ellis_Drossel_time} due to the expansion of the universe \cite{Dod03,Peter_Uzan}) underlies the arrows of time that occur in digital computers and in biology.

\section{Outcomes, How did they get that way, and the key role of values}\label{sec:7}
This paper has discussed how quantum physics processes are shaped by downward causation in both the cases of digital computers and biology, with three key features enabling this. They are, time dependent constraints affecting the potential term in the Hamiltonian, creation and deletion of electrons via quantum field theory effects, and interactions with heat baths playing a key role in wave function collapse and thereby introducing a quantum arrow of time aligned with the cosmological direction of time. \\

\noindent This section emphasizes, 1. The relation to other aspect of quantum physics often discussed, 2. What is the proposal? What have we gained? What have we lost?, 3. How digital computers and humans got to be what they are,  and  4. How the dynamics discussed here regarding digital computers necessarily reflects the purposes and values of the people that create them. 

\subsection{The usual suspects}
This paper has not emphasized some topics common in discussions of quantum biology (for references, see \cite{Ellis_2023}). They are, 
\begin{itemize}
\item \textbf{Quantum tunneling} In biology, there are various cases where tunneling matters \cite{tunnel0,tunnel1,tunnel2,tunnel3}. It is also important in digital computers using tunneling transistors  \cite{tunneling_transistor}. From the perspective of this paper, the key point is that tunneling is always calculated relative to some classical structure, which may be modelled in various ways.
\item \textbf{Superposition} will take place within any  wave-function domain ${\cal U}_\alpha$, as defined in \S 3.2. It will be ended by Contextual Wavefunction Collapse through suitable processes of downward causation  (\S 3.4).
\item \textbf{Entanglement} can also take place within any wave-function domain ${\cal U}_\alpha$, however it will normally rapidly be destroyed by decoherence, for example this will happen in the brain \cite{decoherence}. It is not important in the functioning of ordinary digital computers, however it is central to quantum computing with  quantum entanglement as a computational resource  \cite{Steane}. In this case, the struggle is to maintain  entanglement reliably, which in terms of the present viewpoint is the struggle to maintain wave function domains ${\cal U}_\alpha(t)$ as large as needed, for a substantial time. 
\item \textbf{Quantum spin} is crucial because it underlies the Pauli exclusion principle, and hence is the basis of the periodic table of the elements \cite{Periodic_table0,Periodic_table,Periodic_table1} which is key to existence and structure of digital computers and all biology.
\end{itemize}

\subsection{What is the proposal? What have we gained? What have we lost?} 
If one stands back and asks what the core outcome of this analysis is, it can be briefly summarised as follows. \\

\noindent\textit{What is the proposal?}
\begin{enumerate}
    \item Quantum physics applies everywhere locally.
    \item The crucial feature is that a wave function  $| \psi \rangle_\alpha$ obeying the Schr\"{o}dinger equation (\ref{eq:Schroedinger}) will generally only hold in a local domain  $( {\cal U}_\alpha)(t)$ where the dynamics is lunitary: thus $| \psi \rangle_\alpha  = | \psi \rangle_\alpha ( {\cal U}_\alpha)(t) $.
    \item How big is the domain $( {\cal U}_\alpha)(t)$? The issue is not either physical scale or energy scale: it is whether the dynamics in this domain is linear or not.
\end{enumerate}
\noindent\textit{What have we gained?}
\begin{itemize}
\item The Copenhagen approach: we can assume a classical context within which quantum effects occur locally, 
including effectively classical experimental apparatus; 
\item Contextual Wavefunction Collapse: this classical context will shape quantum outcomes in a downward way;  
\item Classical molecular shape is a result: the basis of molecular biology,
\item A quantum arrow of time, linked to the Direction of Time established by the expanding universe.
\end{itemize}

\noindent \textit{What have we lost?}
\begin{itemize}
    \item 
Schrödinger’s cat: $|\Psi\rangle_{cat}$  does not exist.

\item A single wave function  $|\Psi\rangle_{brain}$ for a brain, 

 hence no many minds interpretation and no “Boltzmann brains”

\item A single wave function  $|\Psi\rangle_{Universe}$ for the entire Universe at all scales, 

hence no Everett interpretation 
\end{itemize}

\subsection{How did they get to be that way?}
The subjects of this discussion (digital computers, life, human beings) are enormously complex. How did they come into being?

\paragraph{Life, and specifically human beings}

Life as we know it came to be what it is by a combination of the processes of (1) \textit{evolution through natural selection}  \cite{Ball_2023,Mitchell_2023,Noble_Noble_2023,Donald}, leading to our genetic inheritance, and (2)  \textit{developmental processes } \cite{Ball_2023,Noble_Noble_2023,Carroll} selectively reading the resultant DNA as needed, which is controlled  by gene regulatory networks \cite{GRN,GRN1,GRN2}. This combination leads to adaptation to the environment: a key example of downward causation \cite{Campbell_1}.

This process of natural selection does not, as some claim, act only at the genetic level \textbf{L5} \cite{Noble_Noble_2023}: on the contrary,  it shapes all emergent levels \textbf{L5} to \textbf{L9} in Table \textbf{2.1} simultaneously so that they work together as needed \cite{Noble_2012}. Thus it selects proteins at level \textbf{L5} for specific biological functions such as enabling vision and transporting oxygen in blood \cite{Wagner}, physiological systems at level \textbf{L8} needed to keep us alive such as the heart pumping blood \cite{Noble_2002} and a brain at Level \textbf{L8} resulting in organisms with agency emerging at Level \textbf{L9} \cite{Mitchell_2023}. As noted above (\S2.2), higher level needs can be met in multiple ways at lower levels, and will reach down to shape configurations of atoms, electrons, and protons at Levels \textbf{L3} and \textbf{L2}  where quantum effects come into play as discussed above. The specific outcomes at these lower levels are shaped by this downward chaining resulting from higher level physiological needs.

\paragraph{Digital computers} All technology also comes into being by similar evolutionary processes of trial and error \cite{Arthur_2010}. The design process is a  careful analysis of needs and how to meet them \cite{design,design1}, aiming to lead to a design that integrates all the emergent levels  \textbf{L4} to \textbf{L9} in \textbf{Table 2.1} so that they jointly lead to the desired operational result through an appropriate computer architecture (\cite{Dasgupta_2016}:\S5, \cite{Mellisinos}:\S2, \cite{Tannen}).
It involves the specific physical shape of the transistors at level \textbf{L5} (\cite{Mellisinos}:\S1), choice of the materials used at level \textbf{L4} and its doping, the design of immensely complex integrated circuits expressing the logic of operation in its design at levels \textbf{L6} to \textbf{L8}. This shapes the computer at all levels by a process of intelligent design: it is ``the science of the artificial''  \cite{Simon_2019}. 

Corresponding to developmental processes in biology are manufacturing processes in technology such as those for transistors \cite{Mellisinos} and integrated circuits \cite{IC}. These again  again have gone through a careful process of adaptive selection: trial and error takes place, discarding those that don't work and improving those that do.  Furthermore factories to produce chips and computers must be designed and created and organised, with technicians who understand the process and know how to handle it. 

Both processes are examples of abstract causation (\S1.2): these explorations results in plans for the computer structure at all levels. Those plans 
are abstract entities that can be realised in many physical forms (in text, on paper, in drawings, in computer files for example). They result from the causal power of human thought, enabled by brains with agency and symbolic understanding that came into being by biological evolutionary processes \cite{Mitchell_2023}. Obviously digital computers would not exist if this were not the case. They are what they are because they were designed and manufactured to be what they are.

\paragraph{Assembly Theory}
In both cases, what can be done at each time depends on what has already happened and what components are available: the principles of Assembly Theory must be fulfilled \cite{assembly,assembly1}. At each moment one takes the next step from what has already been achieved, in a step by step process. Thus  multicellular organisms can only come into being after single cell organisms exist; integrated circuits can only come into being after transistors exist; and so on. Thereby the historical nature of the evolutionary process gets embodied in the physical product. 

\subsection{The key role of values}
The final issue that needs to be emphasized is the key role of values in determining these outcomes. As stated by Dasgupta (\cite{Dasgupta_2016}:31),
\begin{quote}
    ``\textit{All artefacts - engineering and computational - have something in common: they are the products of human thought, human goals, human needs, human desires.} Artefacts are purposive: they reflect the goals of their creators. ... artefacts have entered the world reflecting human needs and goals. \textit{It is not enough to ask what are the laws and principles governing the structure and behaviour of computational artefacts (or for that matter, of pyramids, particle accelerators and kitchen knives) if we then ignore the reason for their existence.''}
\end{quote}
Thus in particular, meaning and purpose underlie computation \cite{Ellis_Drossel_computers}.
The lives of individuals are shaped by their value systems, which constrain and strongly influence their motivation and actions \cite{Noble_Ellis}. Because individuals in turn influence social structures \cite{Berger} which have agency \cite{Elder_Vass}, the  functioning of social structures too is shaped by values to a considerable degree \cite{Ellis_Noble} and this has important real world outcomes \cite{Carney}.

This then goes on to influence the operation of human  creations such as digital computers, as  was emphasized in \cite{Ellis_Drossel_computers}. Computer program and associated algorithms have a set of values associated with them that reflect the purposes and values of the persons writing the program. Their purpose might for example be to write malware intended to infect someone's computer and extract a ransom from them; or it might be to write software used by an international agency for tracking where floods have occurred and relief supplies are needed. The very different set of values in these two cases reflect those of their creators. This is now a key issue because of the way algorithms used in social media have been designed to maximise profits, whatever damage is caused. 

In the end, the sequence of operations at the underlying quantum physics level in digital computers are an expression of a specific set of values; just are the quantum chemistry operations underlying the functioning of our brains.    \\

\textbf{Acknowledgements}: I thank Markus Gabriel and Dean Rickles for useful discussions, and Barbara Drossel for collaborations leading to papers \cite{Ellis_Drossel_computers}, \cite{CWC} and \cite{Ellis_Drossel_time} that are foundational to this project.



\end{document}